\pdfoutput=1
\documentclass[12pt]{article}
\usepackage[utf8]{inputenc} 
\usepackage[T1]{fontenc}
\usepackage{hyperref} 
\usepackage{orcidlink} 
\usepackage{url}      
\usepackage{setspace}
\usepackage{dcolumn}
\usepackage[margin=1in]{geometry}
\usepackage{amsmath}
\usepackage{amssymb} 
\usepackage{mathabx} 
\usepackage{newunicodechar}
\newunicodechar{ξ}{\xi}

\DeclareMathOperator{\sech}{sech} 
\DeclareMathOperator{\csch}{csch} 
\usepackage{graphicx}
\usepackage{subcaption} 
\usepackage{float}
\usepackage{tabularx}
\usepackage[style=numeric-comp, sorting=none]{biblatex} 
\newbibmacro*{pagination}{
}
\bibliography{reference.bib}
\usepackage[mathscr]{eucal}
\usepackage{multirow}
\usepackage{booktabs}
\usepackage[dvipsnames]{xcolor} 
\usepackage{chngcntr} 
\counterwithin{equation}{section} 

\usepackage{xcolor}

\DeclareFieldFormat[article]{title}{\textcolor{magenta}{#1}}

\AtEveryBibitem{%
  \ifentrytype{article}{%
    \renewbibmacro*{journal+issuetitle}{%
      {\color{blue}
        \printfield{journaltitle}%
        \setunit*{\addspace}%
        \printfield{volume}%
        \printfield{number}%
        \setunit{\addspace}%
        \printfield{year}}%
    }%
    \renewbibmacro*{pagination}{%
      \setunit{\addcomma\space}%
      {\color{blue}\printfield{pages}}%
    }%
  }{}%
}

\begin{document}

\title{\textbf{\textcolor{BrickRed}{ {\Large Exploring mutated hilltop inflation in $f(\phi, T)$ gravity with latest CMB and DESI observations}
}}}

\author{ \textcolor{Violet}{\textbf{ Biswajit Deb}} \orcidlink{0000-0001-8992-7600} \footnote{Electronic Address: \textcolor{blue}{biswajit.deb@aus.ac.in}} , \textcolor{Violet}{\textbf{Atri Deshamukhya}} \orcidlink{0000-0003-4350-6645} \footnote{Electronic Address: \textcolor{blue}{atri.deshamukhya@aus.ac.in}}\\
Department of Physics, Assam University, Silchar, India }

\date{}
\maketitle

\begin{abstract}
The high-precision observations from the $Planck$ and BICEP/$Keck$ have imposed stringent bounds on the tensor-to-scalar ratio, ruling out many standard inflationary models. The recent bound from Atacama and SPT along with DESI provides a different viable landscape, making it difficult for the inflationary models to sustain. Amid these increasingly stringent and conflicting observational constraints, modified gravity theories offer a compelling avenue to restore the viability of inflationary models. With this motivation, we study mutated hilltop inflation within $f(\phi, T)$ gravity, where a scalar field is non-minimally coupled to the trace of the energy-momentum tensor. The scalar spectral index $n_s$, the tensor spectral index $n_t$, and the tensor-to-scalar ratio $r$ are calculated, and their trajectories are presented in the $n_s-r$ plane. The results are then confronted with the latest combined CMB bounds from $Planck$, BICEP/$Keck$, ACT DR6, SPT, and DESI DR2. We report that MHI can predict $r \sim 10^{-3}$ and is compatible with the latest CMB \& DESI bound SPA+BK+DESI at $ 1\sigma $ CL.

\end{abstract} \hspace{8pt}


\section{Introduction}	
Inflation is one of the cornerstones of the standard model of cosmology. Since its inception in the early 80's by Alan Guth to solve the horizon and flatness problem associated with the Big Bang model \cite{R1}, the inflationary hypothesis could explain the latest observations to a great extent. For example, it can generate the primordial density fluctuations which later manifest into the large-scale structures of the Universe \cite{R2,R3}  supported by the observational results from WMAP \cite{R4} and $Planck$ \cite{R5}. The simplest model of inflation is achieved by a canonical scalar field called inflaton, which rolls down slowly on its potential giving rise to a de-sitter expansion \cite{R1}. Once the inflation ends, the Universe enters into a radiation-dominated era through a reheating phase \cite{R6,R7}, and particle production begins via leptogenesis, baryogenesis, and nucleosynthesis at different energy scales \cite{R8,R9,R10}. This provides a simple, yet concrete base theory for the early evolution of the Universe.\\ \\
The $\Lambda$CDM model based on General Relativity is the best fit theoretical model till date for describing the Universe \cite{R11}. However, it is yet to explain the origin and smallness of the cosmological constant $\Lambda$ \cite{R12}. Although  GR is successful on the small scales, it has limitations on the large scales. GR cannot explain the dark sector as well as the current accelerated expansion of the Universe \cite{R13,R14}. It fails to explain the cosmic coincidence as well \cite{R15}. These loopholes in GR strongly motivated the cosmologist to look for modifications and alternative theories to Einstein's gravity \cite{R16}.\\ \\
In the 1980's, Starobinsky came out with the $R^2$ inflation model whose result shocked the world \cite{R17}. The Starobinsky model, that is the $f(R)$ gravity model, is not only capable of describing inflation, but it can also explain the late time expansion without the requirement of an exotic energy component in the matter sector \cite{R18,R19}. In fact, $R^2$ inflation is the best-fit inflation model based on the latest result $Planck$ \cite{R20}. Now, modification in the GR action can be done either in the geometry part or in the matter part or in both by adding higher order scalar terms maintaining the Lorentz invariance of the action \cite{R16}. In the last four decades, a plethora of modified theories of gravity emerged in the literature, viz., $f(G)$, $f(Q)$, $f(R,T)$, etc. \cite{R21,R22,R23} which are quite successful in studying astrophysical objects and cosmological phenomena \cite{R24,R25,R32, R26,R27,R28}. Here G, T, and Q stand for the Gauss-Bonnet scalar, trace of the EM tensor and non-metricity, respectively.
\\ \\
Of many such modified theories, Zhang et al. have proposed $f(\phi)T$ gravity, where the non-minimal coupling of the scalar field $\phi$ with the trace of EM tensor $T$ has been added to the gravitational Lagrangian \cite{R29}. This kind of addition of coupling terms is motivated by the quantum gravity framework because it provides a more comprehensive description of gravity unified with other fields which pave the way for the exploration of new physics beyond the scope of general relativity \cite{R30,R31}. The charm of $f(\phi)T$ gravity is that at the end of inflation when the field decays to radiation, Einstein's gravity is recovered naturally. Technically, $f(\phi)T$ gravity is a simple extension of $f(R,T)$ gravity coupled with inflaton. Since $f(R,T)$ gravity presents interesting results for inflation models \cite{R25,R24,R32}, $f(\phi)T$ gravity also holds scope. Zhang et al. studied slow-roll inflation with $f(\phi)T = \sqrt{\kappa}\phi T$ and reported that Chaotic, Natural, and Starobinsky potentials are in better agreement with the data \cite{R29}. Deb et al. reported similar results for the hilltop class of models \cite{R33}. Yeasmin et al. studied warm inflation with the same action and reported that natural potential can be revived \cite{R34}. In addition to this, Ashmita et al. studied other potentials with higher-order coupling terms \cite{R35} and Herrera et al. studied the reheating mechanism in $f(\phi)T$ gravity \cite{R36}. These affirmative results invoke curiosity to investigate leading inflationary models within the $f(\phi)T$ gravity setup. \\ \\
The advanced and precise measurement of CMB anisotropies by $Planck$ ($n_s = 0.9649 \pm 0.0042$) left most theoretical models in doubt \cite{R20}. The latest data release from the Atacama Cosmology Telescope (ACT) and South Pole Telescope (SPT) brought about even more turbulence in the validity of inflationary models \cite{R37,R38,R39}. ACT DR6 provides a larger value of $n_s=0.9666 \pm 0.0077$, ruling out a broad class of inflationary models at the $2\sigma$ confidence level, which includes Starobinsky inflation, Higgs inflation, and $\alpha$-attractors \cite{R38}. Power-law inflation is supported by the ACT data at $1\sigma$ CL \cite{R38}, which was previously ruled out by $Planck$ \cite{R20}. When CMB lensing and Baryon Acoustic Oscillation (BAO) distance measurements from the Dark Energy Spectroscopic Instrument (DESI) are added to $Planck$ and Atacama, the value increases to $n_s = 0.9743 \pm 0.0034$ (P-ACT-LB), which deviates from the original Planck result by $2\sigma$ \cite{R37}. Balkenhol et al. reported a new constraint $n_s=0.9684 \pm 0.0030$ combining $Planck$, ACT, SPT data and named it SPA \cite{R40}. When the B-mode polarization data from the BICEP/$Keck$ telescope is combined with SPA, they reported $n_s=0.9682 \pm 0.0032$ (SPA+BK) which is consistent with the predictions of Starobinsky inflation at $2\sigma$ CL, Higgs inflation at $1.3\sigma$ CL and monomial potentials at $2\sigma$ CL \cite{R40}. Finally, when DESI data are included, the constraint is labeled SPA+BK+DESI and reported as \cite{R40}
\begin{align*}
    n_s &= 0.9728 \pm 0.0029 \\ 
    r &< 0.035
\end{align*}
This constraint is compatible with a range of monomial potentials at $<2\sigma$ CL, Higgs at $2.9\sigma$ CL, however, discard $R^2$ inflation at $3.9\sigma$ CL \cite{R40}. Clearly, when DESI data are added to the CMB, there is a shift in $n_s$ value which puts a strong question on the viability of inflationary models. However, at this point, it is not justified to discard any model on the basis of current observed parameters, unless tension among the data sets is settled. For recent work, references \cite{A1,A2,A3,A4,A5,A6,A7,A8,A9,A10,A11,A12} can be checked.  
\\ \\
In this work, we have taken a generalized form of the coupling term as $f(\phi,T)=\sqrt{\kappa^{4n-3}}\lambda\phi T^n$ which allows to work with any higher order $T$ term corresponding to different values of $n$. Then we considered mutated hilltop inflation, a theoretically motivated model, and investigated how the correction term influences predictions and its viability in light of current observational bounds such as P-ACT-LB and SPA+BK+DESI. Specifically, we will explore two cases $n=1$ and 2 at lower energy scales. \\ \\
The paper is organized as follows. In section 2, we present the formalism of slow-roll inflation in the framework of $f(\phi,T)$ gravity. In section 3, we discuss mutated hilltop inflation with their results in $f(\phi, T)$ gravity and finally, in section 4 we conclude. Throughout the paper, natural units $\hbar=c=\kappa=1$ and (-,+,+,+) sign convention for the metric tensor are used. Where $\kappa=8\pi G=M_{Pl}^{-2}$, $M_{Pl}$ being the reduced $Planck$ mass.

\section{Slow-roll inflation in \texorpdfstring{$f(\phi,T)$}{f(phi,T)} gravity}
The action in $f(\phi,T)$ gravity reads
\begin{equation}
    S= \int d^4x \sqrt{-g} \left[\frac{R}{2\kappa} + f(\phi,T) + L_m \right]
    \label{d1}
\end{equation}
where $R= g^{\alpha\beta}R_{\alpha\beta}$ is the scalar curvature, $f(\phi,T)$ is an arbitrary function of the scalar field $\phi$ non minimally coupled with the trace $T$ of the energy-momentum tensor, $L_m$ is the Lagrangian of the matter sector. \\ \\
To initiate and drive inflation in the early universe, homogeneous scalar fields are required. The canonical scalar field $\phi=\phi(t)$, also known as inflaton, is introduced by the Lagrangian of the form
\begin{equation}
     L_m = -\frac{1}{2}g^{\alpha\beta}\partial_{\alpha}\phi \partial_{\beta}\phi - V(\phi) = \frac{\dot \phi^2}{2} - V(\phi)
     \label{d2}
\end{equation}
where $V(\phi)$ is the potential of the scalar field. Then the energy-momentum tensor $T_{\alpha\beta}$ can be calculated form the matter Lagrangian as
\begin{equation}
    T_{\alpha\beta} = g_{\alpha\beta}L_m - 2 \frac{\delta L_m}{\delta g^{\alpha\beta}} = \partial_{\alpha}\phi \partial_{\beta}\phi + g_{\alpha\beta} \left [ \frac{\dot \phi^2}{2} - V(\phi) \right ]
    \label{d3}
\end{equation}
On metric variation of the action in Eq. \ref{d1}, the modified field equations are obtained as
\begin{equation}
    R_{\alpha\beta} - \frac{1}{2}g_{\alpha\beta}R = \kappa \left[ T_{\alpha\beta} + g_{\alpha\beta} f(\phi,T) - 2 f_T(\phi,T)(T_{\alpha\beta} + \Theta_{\alpha\beta}) \right]
    \label{d4}
\end{equation}
where $\Theta_{\alpha\beta}$ is
\begin{equation}
    \Theta_{\alpha\beta} = g^{\zeta\nu} \frac{\delta T_{\zeta\nu}}{\delta g^{\alpha\beta}} = -2T_{\alpha\beta} + g_{\alpha\beta}L_m - 2 g^{\zeta\nu} \frac{\delta^2 L_m}{\delta g^{\alpha\beta} \delta g^{\zeta\nu}} 
    \label{d5}
\end{equation}
This term $\Theta_{\alpha\beta}$ contains matter Lagrangian $L_m$. So, depending on the type of matter, its form will vary. For this particular study, the form of $f(\phi,T)$ is taken as $f(\phi,T)=\sqrt{\kappa^{4n-3}}\lambda\phi T^n$ where $\lambda$ is a dimensionless model parameter, $n$ is an integer. Since the natural unit system is being used, $\kappa=1$ is set for the rest of the calculation. With this particular choice of $f(\phi,T)$, the field equations are obtained from Eq. \ref{d4} as
\begin{equation}
    R_{\alpha\beta} - \frac{1}{2}g_{\alpha\beta}R = T_{\alpha\beta}^{eff}
    \label{d6}
\end{equation}
where $T_{\alpha\beta}^{eff}$ is the effective energy-momentum tensor containing both the standard matter contribution and the modification induced by the $\phi T^{n}$ coupling term. It is defined as
\begin{equation}
    T_{\alpha\beta}^{eff} =   T_{\alpha\beta} + 2  \lambda \phi T^n  \bigg[ \frac{g_{\alpha\beta}}{2} - \frac{n}{T} ( T_{\alpha\beta} + \Theta_{\alpha\beta}) \bigg] 
    \label{d7}
\end{equation}
It should be noted that the modified field Eqs. \ref{d6} immediately reduce to Einstein field equations for $\lambda=0$. Further, at the end of inflation, when the field is completely decayed to radiation, this model returns to Einstein gravity. Now, from Eq. \ref{d7}, the effective energy density and effective pressure can be obtained as
\begin{equation}
    \rho^{eff} = \frac{\dot \phi^2}{2} + V(\phi) + 2\lambda\phi  \{\dot \phi^{2}-4V(\phi)\}^{n} \left[\frac{n \dot\phi^{2}}{\dot\phi^{2}-4V(\phi)}-\frac{1}{2}\right]
    \label{d8}
\end{equation}
\begin{equation}
    p^{eff} = \frac{\dot \phi^2}{2}  - V(\phi) + \lambda\phi \{ \dot\phi^2-4V(\phi)\}^n 
    \label{d9}
\end{equation}
To study the cosmological implications of the field equations, the Friedmann-Lemaitre-Robertson-Walkar (FLRW) metric was proposed on the basis of cosmological principle. The FLRW metric in the spherical coordinates reads as
\begin{equation}
    ds^2= - dt^2 + a(t)^2 \left[\frac{dr^2}{1-kr^2}+r^2 (d\theta^2 + \sin^2\theta d\phi^2)\right]
    \label{d10}
\end{equation}
where $a(t)$ is the scale factor, $t$ is the cosmic time and $k$ is the spatial curvature with value $\{+1,-1,0\}$ that corresponds to a closed, open and flat Universe, respectively. Since the current observation shows that the present universe is spatially flat, $k=0$ is set for the calculations. Using this metric, the modified Friedmann equations are obtained as
\begin{equation}
    3H^2 =  \frac{\dot \phi^2}{2} + V(\phi) + 2\lambda\phi  \{\dot \phi^{2}-4V(\phi)\}^{n} \left \{\frac{n \dot\phi^{2}}{\dot\phi^{2}-4V(\phi)}-\frac{1}{2} \right \} 
    \label{d11}
\end{equation}
\begin{equation}
    -(2\dot H + 3H^2 ) =  \frac{\dot \phi^2}{2} - V(\phi) + \lambda\phi \{ \dot\phi^2-4V(\phi)\}^n 
    \label{d12}
\end{equation}
where $H=\frac{\dot a(t)}{a(t)}$ is the Hubble parameter. Furthermore, the modified equation of motion can be obtained from Eqs. \ref{d8} and \ref{d9} as
\begin{multline}
    \ddot \phi + V_{\phi}(\phi) + 2 \lambda n \Bigl[ \dot\phi^2 \{\dot\phi^2-4V(\phi)\}^{n-1} + 2\phi \ddot\phi \{ \dot\phi^2-4V(\phi)\}^{n-1} + 2\phi\dot\phi^2(n-1) \{\dot\phi^2-4V(\phi)\}^{n-2}  \\ \{\ddot \phi-2V_{\phi}(\phi)\} \Big] - \lambda \Big[\{\dot\phi^2-4V(\phi)\}^{n}+2n\phi \{\dot\phi^2-4V(\phi)\}^{n-1}\{\ddot \phi-2V_{\phi}(\phi)\} \Big] 
     +3H \dot\phi  \Big[1 + \\ 2 \lambda n \phi \{\dot\phi^2-4V(\phi)\}^{n-1} \Big] = 0 
  \label{d13}
\end{multline}
where $V_{\phi}(\phi) = \frac{dV}{d \phi}$. It should be noted that, under the limit $\lambda=0$,  Eqs. \ref{d11}, \ref{d12} and \ref{d13} reduce to their standard forms in GR. Now, inflation must continue for a prolonged time so it can overcome the horizon problem. For this, the inflaton needs to roll down slowly over its potential for a sufficient time. This requires imposition of slow-roll conditions which are \cite{R29}
\begin{equation}
    \dot \phi^2 \ll V(\phi), \hspace{0.5cm} \ddot \phi \ll 3H\dot\phi, \hspace{0.5cm} \dot\phi^2 \ll H\dot\phi 
    \label{d14}
\end{equation}
Under these slow-roll conditions, the modified Friedmann equation \ref{d11} and the modified equation of motion \ref{d13} become
\begin{equation}
    3H^2 =   V(\phi) -  \lambda \phi \{-4V(\phi)\}^n 
    \label{d15}
\end{equation}
\begin{equation}
     3H\dot \phi \Big[1+2\lambda n \phi \{-4V(\phi)\}^{n-1} \Big] +  V_{\phi}(\phi) -\lambda \Big[\{-4V(\phi)\}^n - 4n\phi V_{\phi}\{-4V(\phi)\}^{n-1} \Big] = 0
     \label{d16}
\end{equation}
Then the modified potential slow-roll parameters are derived from Eqs. \ref{d15} and \ref{d16} as 
\begin{equation}
    \epsilon_V=\frac{\Big[\lambda \big[\{-4V(\phi)\}^n - 4n\phi V_{\phi}(\phi) \{-4V(\phi)\}^{n-1}\big]-V_{\phi}(\phi)   \Big]^2}{2 \Big[1+2\lambda n \phi \{-4V(\phi)\}^{n-1}   \Big]  \Big[V(\phi)- \lambda \phi \{-4V(\phi)\}^n   \Big]^2 }
    \label{d17}
\end{equation}

\begin{multline}
    \eta_V= \frac{1}{ \Big[V(\phi) -\lambda \phi \{-4V(\phi)\}^n \Big] \Big[ 1+2 \lambda n \phi \{-4V(\phi)\}^{n-1} \Big]} \Bigg[ V_{\phi\phi}(\phi)+ 4\lambda n  \Big[2V_{\phi}(\phi) \\  
    \{-4V(\phi)\}^{n-1} + \phi V_{\phi\phi}(\phi) \{-4V(\phi)\}^{n-1} - 4\phi V_{\phi}^{2}(n-1) \{-4V(\phi)\}^{n-2} \Big] + \\  
    \Big[\lambda \Big(\{-4V(\phi)\}^n -4n\phi V_{\phi}(\phi)\{-4V(\phi)\}^{n-1}  \Big)  - V_{\phi}(\phi) \Big] \Big[2\lambda n \Big( \{-4V(\phi)\}^{n-1}  \\ -4\phi V_{\phi}(\phi) (n-1)\{-4V(\phi)\}^{n-2} \Big)  \Big]  \Big[ 1+2 \lambda n \phi \{-4V(\phi)\}^{n-1} \Big]^{-1}  \Bigg]
    \label{d18}
\end{multline} 
It is seen that correction from the term $\phi T^n$ has been induced in the slow-roll parameters, which will play an important role to make the model result consistent with observation. Furthermore, when $\lambda=0$, these slow-roll parameters reduce to their standard GR forms. Inflation continues until $\epsilon_V, |\eta_V| < 1$ and stops when either of the parameters becomes unity \cite{R3}. These slow-roll parameters are crucial to describe the inflation dynamics and its observational imprints on the CMB. \\ \\
The Lagrangian considered in this calculation include the canonical scalar field (Eq. \ref{d2}) and do not contain non-minimal coupling terms of matter with geometry. As a consequence, the field equations can be expressed in a form such that the standard Einstein tensor remains unaltered and the corrections are encapsulated in the effective stress-energy tensor (Eq. \ref{d6} and \ref{d7}). In this case, the speed of sound with which the scalar perturbations travel is trivial and unity \cite{R45}. The modifications from the $ \lambda \phi T^n$ term enter the slow-roll parameters $\epsilon_V$ and $\eta_V$  through the modified Hubble flow. Thus, the standard look of the formulae for the scalar spectral index $n_s$, the tensor spectral index $n_t$, and the tensor-to-scalar ratio $r$ is maintained and the modified gravity effects are encapsulated within the definition of the slow-roll parameters. The expressions for $n_s$, $n_t$, and $r$ read \cite{R3}
\begin{equation}
\begin{split}
    n_s &= 1 - 6 \epsilon_V + 2 \eta_V \\
    n_t &= -2\epsilon_{V} \\
    r &= 16 \epsilon_V     
\end{split}
    \label{d19}
\end{equation}
The amount of inflation is described by the number of $e$-folds $N$ given by
\begin{equation}
    N = \int_{\phi_{end}}^{\phi_{i}} \frac{ \Big[ V(\phi)-\lambda \phi \{-4V(\phi)\}^n \Big]\Big[1+2\lambda n \phi \{-4V(\phi)\}^{n-1}  \Big]}{ \lambda \Big[4n\phi V_{\phi}(\phi) \{-4V(\phi)\}^{n-1}- \{-4V(\phi)\}^n   \Big] + V_{\phi}(\phi)} d\phi
    \label{d20}
\end{equation}
where $\phi_{end}$ is the value of the inflaton at the end of inflation and the upper limit $\phi_{i}$ represents the value of the inflaton at the crossing of the horizon. Thus, any inflation model in $f(\phi,T)$ gravity can now be studied using the potential $V(\phi)$ of that model.


\section{A case study with Mutated Hilltop potential}
Pal et al. proposed a refined version of the hilltop potential, called mutated hilltop potential, which is supergravity inspired and phenomenological \cite{R41}. The form of the potential reads
\begin{equation}
    V(\phi)=V_{0} [ 1- \sech{(\chi \phi)} ]
    \label{d28}
\end{equation}
where $V_0$ is the inflation energy scale and $\chi$ is a parameter with the dimension of $M_{Pl}^{-1}$ that flattens the potential near hilltop, thus creating a sufficient number of $e$-folds. In the mutated hilltop potential, the hyperbolic function with power series expansion includes an infinite number of terms, which makes the model more accurate than the hilltop potentials \cite{R41}. \\ \\
Mutated hilltop potential usually produces a small tensor-to-scalar ratio in general relativity \cite{R41,R42}. It presents viable results in different modified gravity theories as well. Gangopadhyay et al. reported that in Einstein-Gauss-Bonnet gravity, it produces a small $r$ compatible with current observation \cite{R43}. A similar result was reported by Deb et al. in the case of $f(R,T)$ gravity \cite{R32}. Furthermore, in $f(T,\mathcal{T})$ gravity and Rastall-Rainbow gravity, the mutated hilltop potential produces a significantly good result \cite{R44,R46}. Now, what effect does the correction term $\phi T^n$ have on the CMB predictions of the mutated hilltop potentials, will be evaluated in the following section.

\subsection{When n=1}
\begin{figure}[hbt!]
        \centering
        \includegraphics[width=0.55\textwidth]{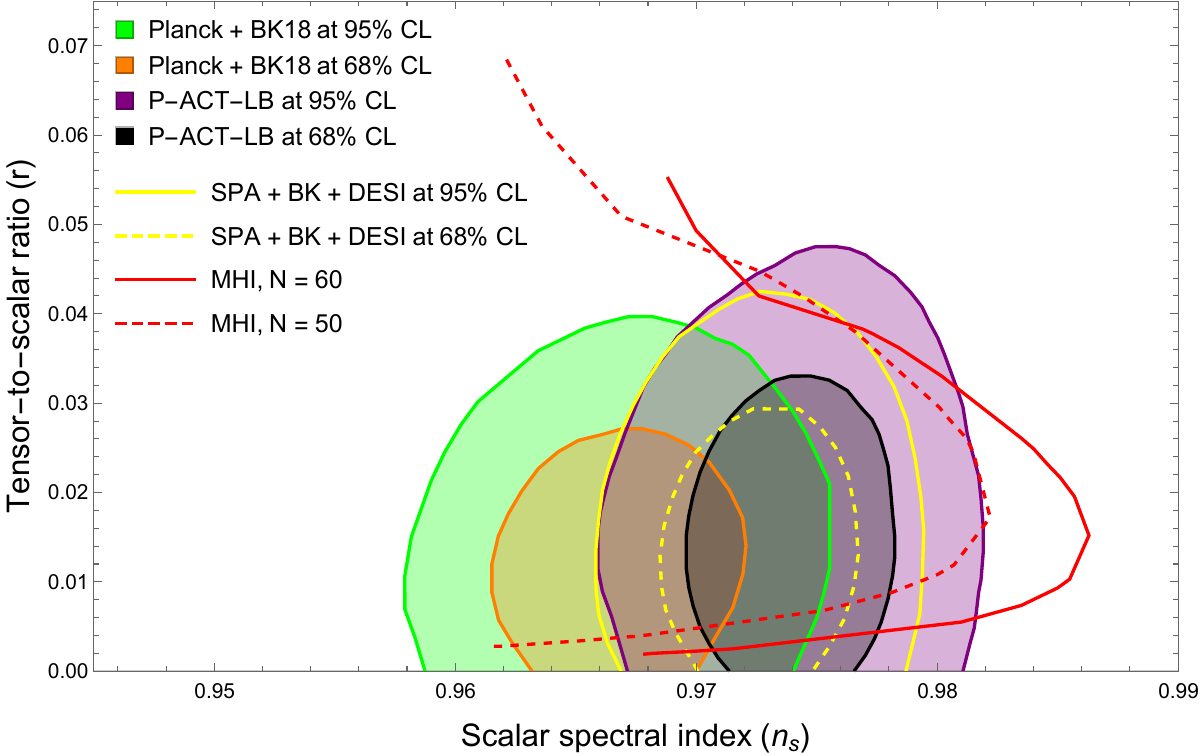}
        \caption {\small The $n_s-r$ trajectory predicted by the mutated hilltop potential for $n=1$ at $\chi=1M_{Pl}^{-1}$ is marked by red solid and dashed line for $N=60$ and $N=50$ $e$-folds respectively. The marginalized joint 68\% and 95\% C.L. regions for $n_s$ and $r$ at $k=0.002 Mpc^{-1}$ from $Planck$+BK18 are shown in orange and green, from P-ACT-LB are shown in black and purple, where as from SPA+BK+DESI are shown by yellow dashed and solid lines respectively.\label{4f4}}
\end{figure}

 When the linear coupling term $\phi T$ is taken i.e., $n=1$, the slow-roll parameters for mutated hilltop potential read as
\begin{equation}
\epsilon_{V}=\frac{
\left[
4\lambda
+\sech(\chi\phi)
\left\{(1+4\lambda\phi)\chi\tanh(\chi\phi)-4\lambda \right\}
\right]^{2}
}{
2(1+2\lambda\phi)(1+4\lambda\phi)^{2}
\{\sech(\chi\phi)-1\}^{2}
} \label{d29}
\end{equation}

\begin{multline}
\eta_{V}=\frac{1}{2(1+2\lambda\phi)^{2}(1+4\lambda\phi)\{\sech(\chi\phi)-1\}
}\Big[
\sech^{3}(\chi\phi)
\Big\{
-3\chi^{2}-8\lambda^{2}-18\chi^{2}\lambda\phi
\\ -24\chi^{2}\lambda^{2}\phi^{2}
+12\lambda^{2}\cosh(\chi\phi)
+[\chi^{2}(1+6\lambda\phi+8\lambda^{2}\phi^{2})-8\lambda^{2}] \cosh(2\chi\phi)
\\
+4\lambda^{2}\cosh(3\chi\phi)
-6\chi\lambda\sinh(2\chi\phi)
-8\chi\lambda^{2}\phi\sinh(2\chi\phi)
\Big\}\Big] \label{d30}
\end{multline}
Using Eqs. \ref{d19}, \ref{d29} and \ref{d30}, the tensor-to-scalar ratio and scalar spectral index are calculated for $N=50$ and 60 $e$-folds by varying the model parameter $\lambda$ and setting $\chi$ at $1M_{Pl}^{-1}$. Then the result is plotted in the $n_s-r$ plane, presented in Figure \ref{4f4}. It is found that both $n_s$ and $r$ are extremely sensitive to the parameter $\lambda$. The trajectories for $N=50$ and 60 $e$-folds fall into the $1\sigma$ region of $Planck$+BK18, P-ACT-LB, and the most recent CMB bound from SPA+BK+DESI. With a suitable choice of $\lambda$, the tensor-to-scalar ratio can be achieved at the order of $10^{-3}$, which will remain compatible with the upcoming CMB bounds such as LiteBIRD \cite{R47} and CMB-S4 \cite{R48}. Furthermore, the model parameter space is constrained considering $Planck$+BK18 bound which is $0.002 > \lambda > 10^{-5}$ for $N=50$ and $0.0004 > \lambda > 10^{-5}$ for $N=60$ respectively. 

\subsection{When n=2}
In case of the quadratic trace term coupled with the inflaton, the slow-roll parameters for mutated hilltop potential read
\begin{multline}
    \epsilon_{V}=\frac{1}{2\{\sech(\chi\phi)-1\}^{2}\{1-16\lambda V_{0}\phi+16\lambda V_{0}\sech(\chi\phi)\}^{3}}\Big[-16\lambda V_{0}+16\lambda V_{0}\sech^2(\chi\phi)\\ \{2\chi\phi\tanh(\chi\phi)-1 \}
    +\sech(\chi\phi)\{32\lambda V_{0}+(1-32\lambda V_{0}\phi)\chi\tanh(\chi\phi)\}\Big]^{2} \label{d31}
\end{multline}

\begin{multline}
\eta_{V}=
\frac{\cosh^{3}(\chi\phi)\,\csch^{2}\!\left(\frac{\chi\phi}{2}\right)}
{4\left[-16\lambda V_{0}\phi+(-1+16\lambda V_{0}\phi)\cosh(\chi\phi)\right]^{2}}
\times \Bigg[
-\chi^{2}\left(-3+\cosh(2\chi\phi)\right)\sech^{3}(\chi\phi)
\\
+64\,\chi\,\lambda V_{0}\,\sech^{4}(\chi\phi)\,
\sinh^{2}\!\left(\frac{\chi\phi}{2}\right)
\left[-5\chi\phi-2\chi\phi\cosh(\chi\phi)
+\chi\phi\cosh(2\chi\phi)-2\sinh(2\chi\phi)\right]
\\
+\frac{32\lambda V_{0}\left[1+\sech(\chi\phi)\left(-1+\chi\phi\tanh(\chi\phi)\right)\right]}
{1-16\lambda V_{0}\phi+16\lambda V_{0}\phi\,\sech(\chi\phi)}
\times
\Big[
-16\lambda V_{0}
+16\lambda V_{0}\sech^{2}(\chi\phi)
\left(-1+2\chi\phi\tanh(\chi\phi)\right)
\\
+\sech(\chi\phi)
\left(32\lambda V_{0}+(\chi-32\chi\lambda V_{0}\phi)\tanh(\chi\phi)\right)
\Big]
\Bigg] \label{d32}
\end{multline}

\begin{figure}[hbt!]
        \centering
        \includegraphics[width=0.55\textwidth]{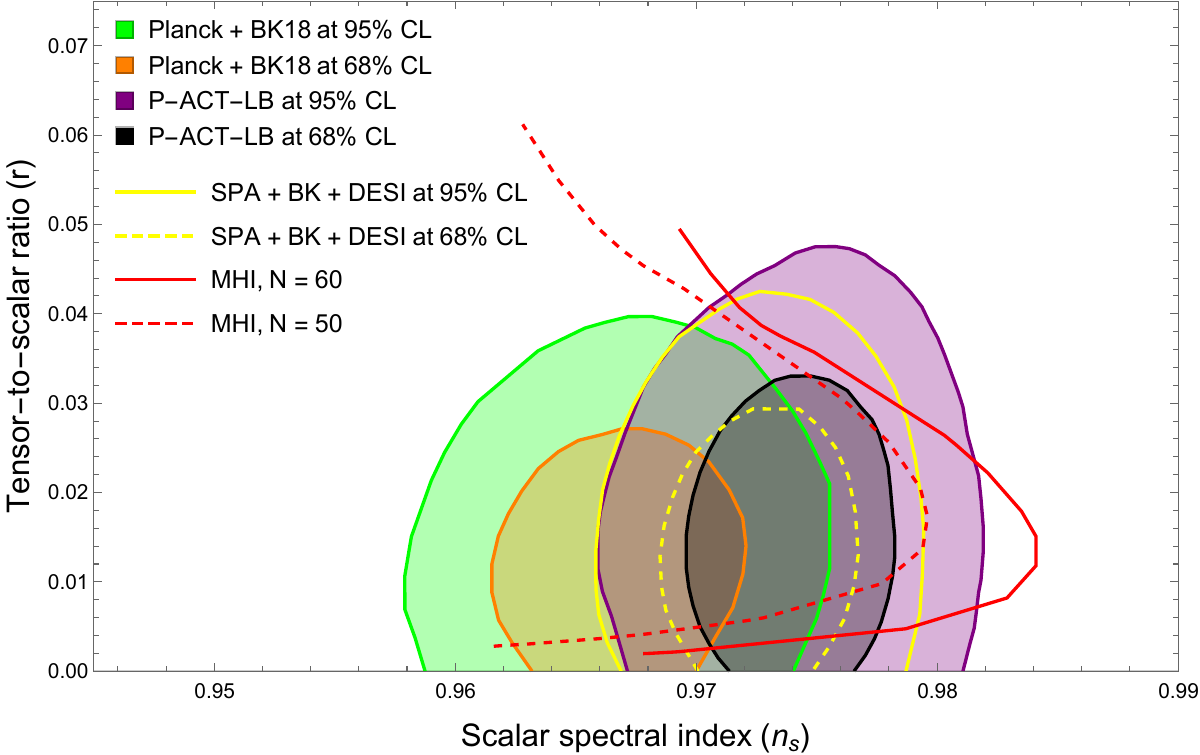}
        \caption {\small The $n_s-r$ trajectory predicted by the mutated hilltop potential for $n=2$ at $\chi=1M_{Pl}^{-1}$ is marked by red solid and dashed line for $N=60$ and $N=50$ $e$-folds respectively. The marginalized joint 68\% and 95\% C.L. regions for $n_s$ and $r$ at $k=0.002 Mpc^{-1}$ from $Planck$+BK18 are shown in orange and green, from P-ACT-LB are shown in black and purple, where as from SPA+BK+DESI are shown by yellow dashed and solid lines respectively.\label{4f5}}
\end{figure}

Then using Eqs. \ref{d19}, \ref{d31} and \ref{d32}, the tensor-to-scalar ratio and scalar spectral index are calculated for different values of $\lambda$. It is found that positive values of $\lambda$ do not work in this case, however,  negative $\lambda$ can make the $n_s-r$ predictions of the mutated hilltop potential aligned to the observed CMB bounds. At $\chi=1M_{Pl}^{-1}$, the trajectories for $N=50$ and 60 $e$-folds are presented in Figure \ref{4f5}. Both trajectories cover entire $1\sigma$ regions of the $Planck$+BK18, P-ACT-LB, and SPA+BK+DESI data. It is seen that $n_s$ and $r$ are highly sensitive to the parameter $\lambda$ and $r \sim 10^{-3}$ can be achieved for certain values of $\lambda$. Then the parameter space of $\lambda$ is evaluated considering $Planck$+BK18 bound which is found to be $-0.05 < \lambda < -10^{-4}$ for $N=50$ and  $-0.009 < \lambda < - 10^{-4}$ for $N=60$ respectively. 

\subsection{Sensitivity analysis of the potential parameter $\chi$} 
\begin{figure}[h!]
 \centering
 \begin{subfigure}[h]{0.49\textwidth}
     \centering
     \includegraphics[width=\textwidth]{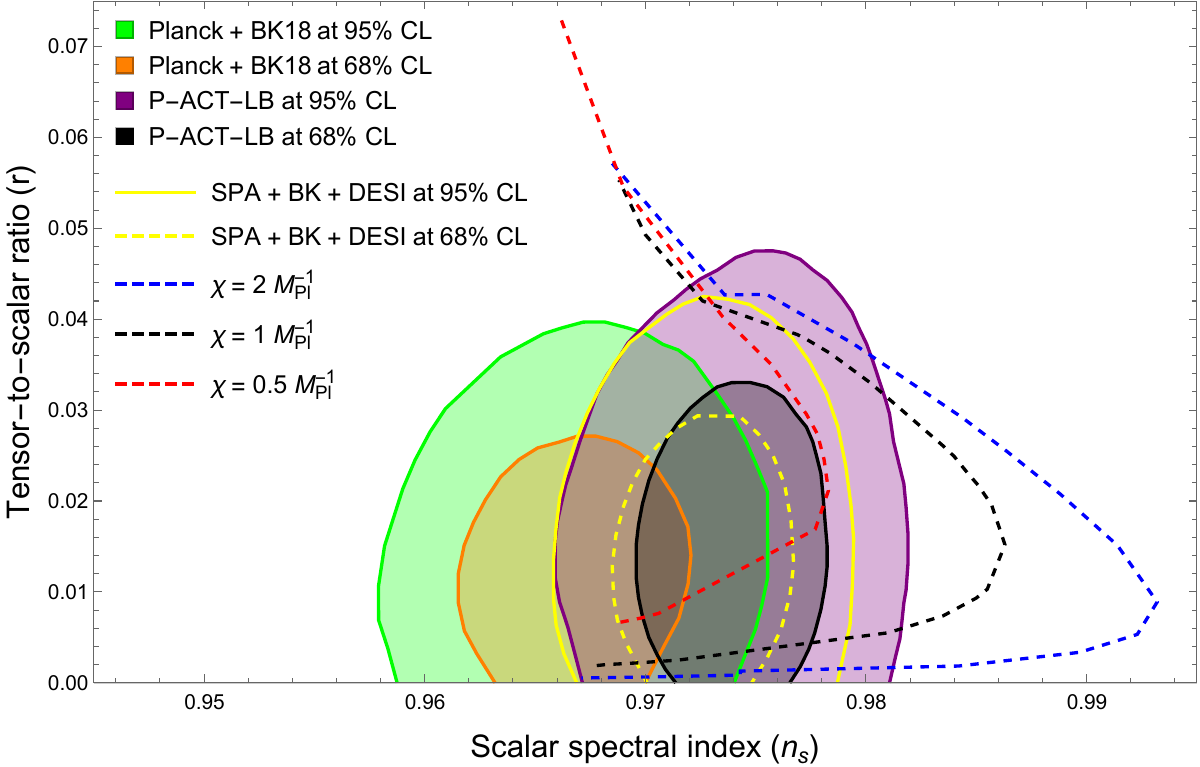}
     \caption{$n=1$}
 \end{subfigure}
 \hfill
 \begin{subfigure}[h]{0.49\textwidth}
     \centering
     \includegraphics[width=\textwidth]{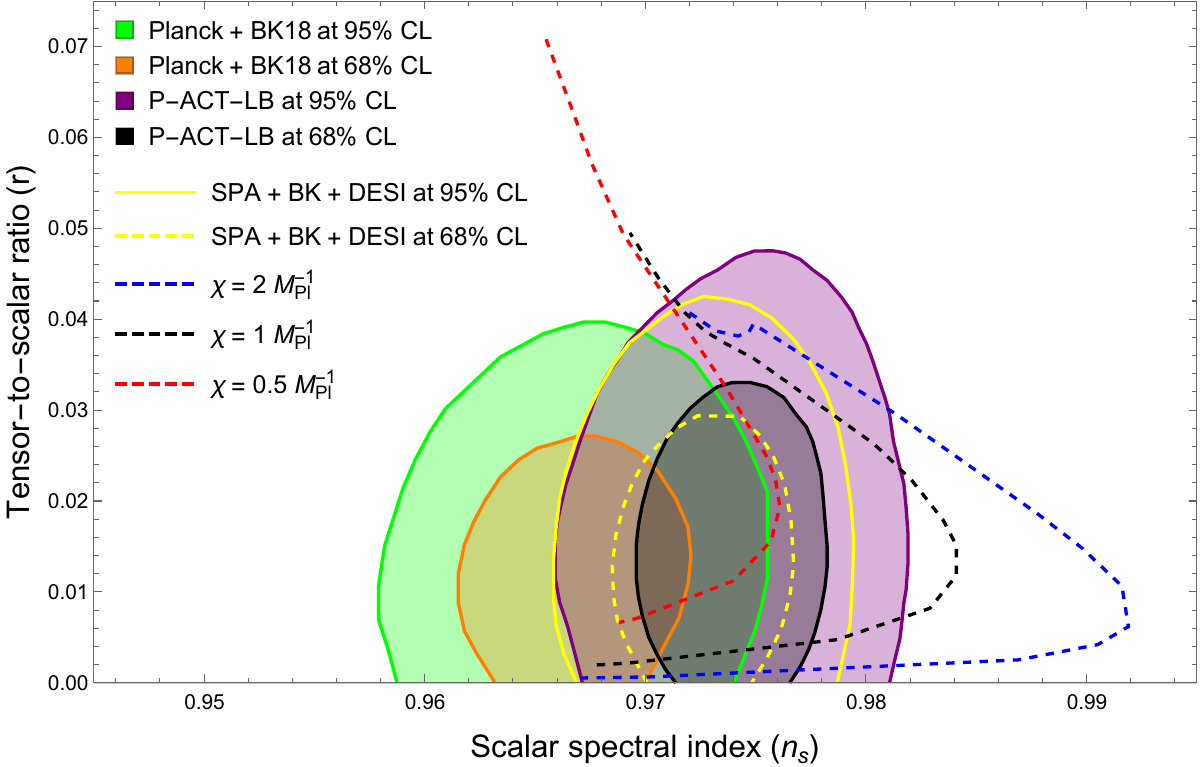}
     \caption{$n=2$}
 \end{subfigure}
 \caption{ The $n_s-r$ trajectories for  $\chi= 2M_{Pl}^{-1}$, $\chi= 1 M_{Pl}^{-1}$ and $\chi= 0.5M_{Pl}^{-1}$ marked by the dashed blue, black, and red colour respectively for $N=60$ $e$-folds. The left figure is for $n=1$ case, whereas, the right one stands for $n=2$. The marginalized joint 68\% and 95\% C.L. regions for $n_s$ and $r$ at $k=0.002 Mpc^{-1}$ from $Planck$+BK18 are shown in orange and green, from P-ACT-LB are shown in black and purple, where as from SPA+BK+DESI are shown by yellow dashed and solid lines respectively. }
 \label{f3}   
 \end{figure}

The potential parameter $\chi$ controls the steepness and flatness of the potential. It sets the characteristic field scale over which the potential varies and tells how sensitive the potential is to changes in the inflaton field $\phi$. For small values of $\chi$, the potential becomes wider and flatter around the hilltop, leading to a slower rolling of the inflaton and prolonged inflation. However, a large value of $\chi$ makes the potential steeper, resulting in a faster evolution with fewer $e$-folds. Therefore, with the change in $\chi$, the inflationary predictions will change quantitatively while the overall shape of the potential remains preserved. \\ \\
Now, to examine the robustness of the previous results for the choice of the potential parameter $\chi$, a sensitivity analysis is performed considering three values, viz., $\chi=0.5,1$, and $2 M_{Pl}^{-1}$, respectively. In Figure \ref{f3}, the trajectories in the $n_s-r$ plane are presented for the three choices of $\chi$ at $N=60 $ $e$-folds. It is observed that for both the cases with $n=1$ and $n=2$, with the increase in $\chi$ the predicted inflationary trajectory shifts towards larger values of $n_s$, whereas the tensor-to-scalar ratio decreases moderately. Therefore, smaller values $\chi$ are favorable. Nevertheless, for all three considered values of $\chi$, the model remains compatible with the current observational constraints at $1\sigma$  CL, for suitable choices of the coupling parameter $\lambda$. The corresponding ranges of the coupling parameter $\lambda$ (evaluated for $Planck$+BK18) are listed in the table \ref{T1}.

\begin{table}
    \centering
    {
    \begin{tabular}{ccc} \toprule
      Case   & $\chi$ (in $M_{Pl}^{-1}$)  & Range of $\lambda$ \\ \midrule
         & 0.5 & $0.0002 > \lambda > 10^{-5}$ \\ 
       $n=1$  & 1 & $0.0004 > \lambda > 10^{-5}$\\
         & 2 & $0.0003 > \lambda > 10^{-5}$ \\ \midrule
         & 0.5 & $ - 0.005 < \lambda < - 10^{-4}$ \\
       $n=2$  & 1 & $ - 0.009 < \lambda < - 10^{-4}$ \\
         & 2 & $ - 0.007 < \lambda < - 10^{-4}$ \\ \bottomrule
    \end{tabular}
    \caption{Allowed range for $\lambda$ considering $Planck$+BK18 for different choice of $\chi$.}}
    \label{T1}
    
\end{table} 

From table \ref{T1}, it is evident that with the change in $\chi$, a slight shift in the numerical range of the allowed coupling parameter $\lambda$ is observed, however, the overall order of magnitude remains same. It indicates that the observables $n_s$ and $r$ are more sensitive to the modified gravity parameter $\lambda$ than to subtle variations of the potential parameter $\chi$.

\subsection{Comparison of MHI with hilltop inflation}
The general form of the hilltop potential reads
\begin{equation}
   V(\phi)=V_{0} \left[1-\left(\frac{\phi}{\mu}\right)^m\right]
\end{equation}
where $V_0$ is the inflationary scale, $\mu$ is the vacuum expectation value ($vev$) of the inflaton and $m$ is a positive integer whose different value correspond to different hilltop models. The quartic model $(m=4)$ in GR is in good agreement with the $Planck$ 18, however, $\mu \gg 10 M_p$ is required \cite{R20}. Now in GR, mutated hilltop inflation predicts lower tensor-to-scalar ratio in comparison to standard hilltop inflation \cite{R41}. A similar trend is observed in case of $f(\phi,T)$ gravity. Deb et. al showed that the trajectories of different hilltop potentials in $f(\phi,T)$ gravity cover only $2\sigma$ regions of $Planck$ \cite{R33}. However, from Figure \ref{4f4}, and \ref{4f5} it is clear that the MHI trajectory covers the $1\sigma$ region of advanced SPA+BK+DESI bound, hinting at the superiority of MHI over HI. For a quantitative comparison between these two models, their predictions are presented in the Table \ref{tab:sensitivity}.

\begin{table*}[ht]
\centering
{
\begin{tabular}{|c|c|c|c|c|c|c|}
\hline
$n$ & Inflationary Model & Fixed Parameter & $\lambda$ & $N$ & $n_s$ & $r$ \\
\hline

\multirow{6}{*}{1}
& \multirow{2}{*}{Hilltop ($m=2$)}
& \multirow{2}{*}{$\mu=5M_{\rm Pl}$}
& \multirow{2}{*}{5}
& 60 & 0.9715 & 0.09 \\ \cline{5-7}
&&&& 50 & 0.9661 & 0.10 \\ \cline{2-7}

& \multirow{2}{*}{Hilltop ($m=4$)}
& \multirow{2}{*}{$\mu=5M_{\rm Pl}$}
& \multirow{2}{*}{60}
& 60 & 0.9737 & 0.08 \\ \cline{5-7}
&&&& 50 & 0.9686 & 0.09 \\ \cline{2-7}

& \multirow{2}{*}{MHI}
& \multirow{2}{*}{$\chi=1M_{\rm Pl}^{-1}$}
& \multirow{2}{*}{$10^{-4}$}
& 60 & 0.9682 & 0.002 \\ \cline{5-7}
&&&& 50 & 0.9619 & 0.002 \\ \hline

\multirow{6}{*}{2}
& \multirow{2}{*}{Hilltop ($m=2$)}
& \multirow{2}{*}{$\mu=5M_{\rm Pl}$}
& \multirow{2}{*}{10}
& 60 & 0.9705 & 0.11 \\ \cline{5-7}
&&&& 50 & 0.9649 & 0.13 \\ \cline{2-7}

& \multirow{2}{*}{Hilltop ($m=4$)}
& \multirow{2}{*}{$\mu=5M_{\rm Pl}$}
& \multirow{2}{*}{10}
& 60 & 0.9539 & 0.24 \\ \cline{5-7}
&&&& 50 & 0.9458 & 0.29 \\ \cline{2-7}

& \multirow{2}{*}{MHI}
& \multirow{2}{*}{$\chi=1M_{\rm Pl}^{-1}$}
& \multirow{2}{*}{$-10^{-3}$}
& 60 & 0.9680 & 0.001 \\ \cline{5-7}
&&&& 50 & 0.9617 & 0.002 \\ \hline

\end{tabular}
}
\caption{Predicted values of $n_s$ and $r$ for MHI and HI in $f(\phi,T)$ gravity.}
\label{tab:sensitivity}

\end{table*}
The comparison clearly demonstrates that both models predict the scalar spectral index $n_s$ consistent with current observational constraints. However, they differ in their tensor-to-scalar ratio significantly. MHI predicts tensor-to-scalar ratio to the order of $10^{-3}$ whereas the hilltop models predict much higher values of $r$ compared to the current upper bound from the $Planck$+BK18 ($r<0.036$). Consequently, MHI is not only compatible with the current observational bound but also with the future expected bounds such as the LiteBIRD and the CMB-S4.

\section{Conclusion}
Modified theories of gravity are at the forefront of current theoretical research. It bridges the early and late-time evolutions and simultaneously explains astrophysical compact objects. Having this motivation, we started with the aim of studying mutated hilltop inflation within $f(\phi,T)$ gravity, an extended version of $f(R,T)$ gravity. We choose the simplest yet generalized coupling term $f(\phi,T)=\sqrt{\kappa^{4n-3}}\lambda\phi T^n$ in the gravitational Lagrangian with special emphasis on two cases $n=1,2$ and studied its effects on the cosmological observables viz. the scalar spectral index $n_s$, the tensor spectral index $n_t$, and the tensor-to-scalar ratio $r$. Furthermore, the model predictions are presented in the $n_s-r$ plane and confronted with the latest bounds from the CMB probes such as $Planck$, BICEP/$Keck$, Atacama, South Pole Telescope, and DESI. The results are summarized as follows:
\begin{itemize}
    \item \textbf{Mutated hilltop potential with $n=1$} \\
    MHI is found to retain its excellent results in terms of viability with observations in $f(\phi, T)$ gravity. Within a suitable parameter space of $\lambda$, the $n_s-r$ curves for $N=50$ and 60 $e$-folds not only satisfy $Planck$+BK18 at $1\sigma$ CL, but also agree with the elevated $n_s$ bounds provided by the Atacama and DESI probes. Furthermore, it satisfies the combined bound SPA+BK+DESI at $1\sigma$ CL. The tensor-to-scalar ratio can be suppressed to the order of $10^{-3}$ by fine tuning $\lambda$, and the tensor spectral index is found to be of $n_t \sim - 10^{-4}$. Finally, the parameter space of the model is limited considering the $Planck$+BK18 bound which is $0.002 > \lambda > 10^{-5}$ for $N=50$ and $0.0004 > \lambda > 10^{-5}$ for $N=60$, respectively. 

    \item \textbf{Mutated hilltop potential with $n=2$} \\
     When the $\phi T^2$ correction is applied to MHI, it is observed that the model predictions are sensitive to the negative values of the coupling parameter only. The $n_s-r$ trajectories fall into the $1\sigma$ regions of all CMB bounds of different probes. The tensor-to-scalar ratio is seen to be slightly elevated in the case of $n=2$ compared to the case of $n=1$. However, with a suitable choice of $\lambda$, $r \sim 10^{-3}$ is achievable, making it compatible with LiteBIRD and CMB-S4. The tensor spectral index is also in agreement with the observed results. The model parameter space is reported as $-0.05 < \lambda < -10^{-4}$ for $N=50$ and  $-0.009 < \lambda < - 10^{-4}$ for $N=60$ respectively, considering $Planck$+BK18 bounds. 

\end{itemize}
Clearly, the non-minimal coupling of inflaton with the trace of the EM tensor affects positively mutated hilltop potential, allowing it to be consistent with future CMB probes like LiteBIRD and CMB-S4. The sensitivity analysis of the potential parameter $\chi$ indicates that the model predictions are more sensitive to the coupling parameter $\lambda$ than to the $\chi$. Furthermore, it is also noted that MHI predictions are in better agreement with the current bound than the standard HI models in $f(\phi,T)$ gravity. The analysis has also been extended for the $n=-1$ case and found that MHI does not work well with this correction term.\\ \\
In this work, the trace of the EM tensor is calculated only from the canonical scalar field. However, it can be sourced from non-canonical scalar fields as well as from multiple scalar fields. This theoretical setup might present interesting results which we leave as a scope for future work.


\printbibliography

@article{R1,
    author = "Guth, Alan H.",
    editor = "Fang, Li-Zhi and Ruffini, R.",
    title = "{The Inflationary Universe: A Possible Solution to the Horizon and Flatness Problems}",
    reportNumber = "SLAC-PUB-2576",
    doi = "10.1103/PhysRevD.23.347",
    journal = "Phys. Rev. D",
    volume = "23",
    pages = "347--356",
    year = "1981"
}

@article{R2,
    author = "Kofman, L. A. and Linde, Andrei D.",
    title = "{Generation of Density Perturbations in the Inflationary Cosmology}",
    reportNumber = "TARTU-A-4-1986",
    doi = "10.1016/0550-3213(87)90698-5",
    journal = "Nucl. Phys. B",
    volume = "282",
    pages = "555",
    year = "1987"
}

@book{R3,
  title={Cosmological inflation and large-scale structure},
  author={Liddle, Andrew R and Lyth, David H},
  year={2000},
  publisher={Cambridge university press}
}

@article{R4,
    author = "Hinshaw, G. and others",
    collaboration = "WMAP",
    title = "{Nine-Year Wilkinson Microwave Anisotropy Probe (WMAP) Observations: Cosmological Parameter Results}",
    eprint = "1212.5226",
    archivePrefix = "arXiv",
    primaryClass = "astro-ph.CO",
    doi = "10.1088/0067-0049/208/2/19",
    journal = "Astrophys. J. Suppl.",
    volume = "208",
    pages = "19",
    year = "2013"
}

@article{R5,
    author = "Aghanim, N. and others",
    collaboration = "Planck",
    title = "{Planck 2018 results. VI. Cosmological parameters}",
    eprint = "1807.06209",
    archivePrefix = "arXiv",
    primaryClass = "astro-ph.CO",
    doi = "10.1051/0004-6361/201833910",
    journal = "Astron. Astrophys.",
    volume = "641",
    pages = "A6",
    year = "2020",
    note = "[Erratum: Astron.Astrophys. 652, C4 (2021)]"
}

@article{R6,
    author = "Shtanov, Y. and Traschen, Jennie H. and Brandenberger, Robert H.",
    title = "{Universe reheating after inflation}",
    eprint = "hep-ph/9407247",
    archivePrefix = "arXiv",
    reportNumber = "BROWN-HET-957",
    doi = "10.1103/PhysRevD.51.5438",
    journal = "Phys. Rev. D",
    volume = "51",
    pages = "5438--5455",
    year = "1995"
}

@article{R7,
    author = "Bassett, Bruce A. and Tsujikawa, Shinji and Wands, David",
    title = "{Inflation dynamics and reheating}",
    eprint = "astro-ph/0507632",
    archivePrefix = "arXiv",
    doi = "10.1103/RevModPhys.78.537",
    journal = "Rev. Mod. Phys.",
    volume = "78",
    pages = "537--589",
    year = "2006"
}

@article{R8,
    author = "Davidson, Sacha and Nardi, Enrico and Nir, Yosef",
    title = "{Leptogenesis}",
    eprint = "0802.2962",
    archivePrefix = "arXiv",
    primaryClass = "hep-ph",
    doi = "10.1016/j.physrep.2008.06.002",
    journal = "Phys. Rept.",
    volume = "466",
    pages = "105--177",
    year = "2008"
}

@article{R9,
  title={Big bang nucleosynthesis-Theories and observations},
  author={Boesgaard, Ann Merchant and Steigman, Gary},
  journal={IN: Annual review of astronomy and astrophysics. Volume 23 (A86-14507 04-90). Palo Alto, CA, Annual Reviews, Inc., 1985, p. 319-378.},
  volume={23},
  pages={319--378},
  year={1985}
}

@article{R10,
    author = "Olive, K. A. and others",
    collaboration = "Particle Data Group",
    title = "{Review of Particle Physics}",
    doi = "10.1088/1674-1137/38/9/090001",
    journal = "Chin. Phys. C",
    volume = "38",
    pages = "090001",
    year = "2014"
}

@article{R11,
    author = "Ooba, Junpei and Ratra, Bharat and Sugiyama, Naoshi",
    title = "{Planck 2015 Constraints on the Non-flat $\Lambda$CDM Inflation Model}",
    eprint = "1707.03452",
    archivePrefix = "arXiv",
    primaryClass = "astro-ph.CO",
    doi = "10.3847/1538-4357/aad633",
    journal = "Astrophys. J.",
    volume = "864",
    number = "1",
    pages = "80",
    year = "2018"
}

@article{R12,
    author = "Del Popolo, Antonino and Le Delliou, Morgan",
    title = "{Small scale problems of the $\Lambda$CDM model: a short review}",
    eprint = "1606.07790",
    archivePrefix = "arXiv",
    primaryClass = "astro-ph.CO",
    doi = "10.3390/galaxies5010017",
    journal = "Galaxies",
    volume = "5",
    number = "1",
    pages = "17",
    year = "2017"
}

@article{R13,
    author = "Bertone, Gianfranco and Hooper, Dan",
    title = "{History of dark matter}",
    eprint = "1605.04909",
    archivePrefix = "arXiv",
    primaryClass = "astro-ph.CO",
    reportNumber = "FERMILAB-PUB-16-157-A",
    doi = "10.1103/RevModPhys.90.045002",
    journal = "Rev. Mod. Phys.",
    volume = "90",
    number = "4",
    pages = "045002",
    year = "2018"
}

@article{R14,
    author = "Frieman, Joshua and Turner, Michael and Huterer, Dragan",
    title = "{Dark Energy and the Accelerating Universe}",
    eprint = "0803.0982",
    archivePrefix = "arXiv",
    primaryClass = "astro-ph",
    reportNumber = "FERMILAB-PUB-08-613-A",
    doi = "10.1146/annurev.astro.46.060407.145243",
    journal = "Ann. Rev. Astron. Astrophys.",
    volume = "46",
    pages = "385--432",
    year = "2008"
}

@article{R15,
    author = "Velten, H. E. S. and vom Marttens, R. F. and Zimdahl, W.",
    title = "{Aspects of the cosmological {\textquotedblleft}coincidence problem{\textquotedblright}}",
    eprint = "1410.2509",
    archivePrefix = "arXiv",
    primaryClass = "astro-ph.CO",
    doi = "10.1140/epjc/s10052-014-3160-4",
    journal = "Eur. Phys. J. C",
    volume = "74",
    number = "11",
    pages = "3160",
    year = "2014"
}

@article{R16,
    author = "Shankaranarayanan, S. and Johnson, Joseph P.",
    title = "{Modified theories of gravity: Why, how and what?}",
    eprint = "2204.06533",
    archivePrefix = "arXiv",
    primaryClass = "gr-qc",
    doi = "10.1007/s10714-022-02927-2",
    journal = "Gen. Rel. Grav.",
    volume = "54",
    number = "5",
    pages = "44",
    year = "2022"
}

@article{R17,
    author = "Starobinsky, Alexei A.",
    editor = "Khalatnikov, I. M. and Mineev, V. P.",
    title = "{A New Type of Isotropic Cosmological Models Without Singularity}",
    doi = "10.1016/0370-2693(80)90670-X",
    journal = "Phys. Lett. B",
    volume = "91",
    pages = "99--102",
    year = "1980"
}

@article{R18,
    author = "Odintsov, S. D. and Oikonomou, V. K.",
    title = "{Unification of Inflation with Dark Energy in $f(R)$ Gravity and Axion Dark Matter}",
    eprint = "1905.03496",
    archivePrefix = "arXiv",
    primaryClass = "gr-qc",
    doi = "10.1103/PhysRevD.99.104070",
    journal = "Phys. Rev. D",
    volume = "99",
    number = "10",
    pages = "104070",
    year = "2019"
}

@article{R19,
    author = "Nojiri, Shin'ichi and Odintsov, Sergei D. and Oikonomou, V. K.",
    title = "{Unifying Inflation with Early and Late-time Dark Energy in $F(R)$ Gravity}",
    eprint = "1912.13128",
    archivePrefix = "arXiv",
    primaryClass = "gr-qc",
    doi = "10.1016/j.dark.2020.100602",
    journal = "Phys. Dark Univ.",
    volume = "29",
    pages = "100602",
    year = "2020"
}

@article{R20,
    author = "Akrami, Y. and others",
    collaboration = "Planck",
    title = "{Planck 2018 results. X. Constraints on inflation}",
    eprint = "1807.06211",
    archivePrefix = "arXiv",
    primaryClass = "astro-ph.CO",
    doi = "10.1051/0004-6361/201833887",
    journal = "Astron. Astrophys.",
    volume = "641",
    pages = "A10",
    year = "2020"
}

@article{R21,
    author = "Nojiri, Shin'ichi and Odintsov, Sergei D.",
    title = "{Modified Gauss-Bonnet theory as gravitational alternative for dark energy}",
    eprint = "hep-th/0508049",
    archivePrefix = "arXiv",
    doi = "10.1016/j.physletb.2005.10.010",
    journal = "Phys. Lett. B",
    volume = "631",
    pages = "1--6",
    year = "2005"
}

@article{R22,
    author = "Beltr{\'a}n Jim{\'e}nez, Jose and Heisenberg, Lavinia and Koivisto, Tomi",
    title = "{Coincident General Relativity}",
    eprint = "1710.03116",
    archivePrefix = "arXiv",
    primaryClass = "gr-qc",
    reportNumber = "NORDITA-2017-100, IFT-UAM/CSIC-17-093, ITS-ETH-2017-10",
    doi = "10.1103/PhysRevD.98.044048",
    journal = "Phys. Rev. D",
    volume = "98",
    number = "4",
    pages = "044048",
    year = "2018"
}

@article{R23,
    author = "Harko, Tiberiu and Lobo, Francisco S. N. and Nojiri, Shin'ichi and Odintsov, Sergei D.",
    title = "{$f(R,T)$ gravity}",
    eprint = "1104.2669",
    archivePrefix = "arXiv",
    primaryClass = "gr-qc",
    doi = "10.1103/PhysRevD.84.024020",
    journal = "Phys. Rev. D",
    volume = "84",
    pages = "024020",
    year = "2011"
}

@article{R24,
    author = "Yeasmin, Sabina and Deb, Biswajit and Deshamukhya, Atri",
    title = "{Warm inflation in $f(R,T)$ gravity}",
    eprint = "2211.05059",
    archivePrefix = "arXiv",
    primaryClass = "gr-qc",
    doi = "10.1016/j.cjph.2023.07.004",
    journal = "Chin. J. Phys.",
    volume = "85",
    pages = "359--374",
    year = "2023"
}

@article{R25,
    author = "Deb, Biswajit and Deshamukhya, Atri",
    title = "{Inflation in $f(R,T)$ gravity with double-well potential}",
    eprint = "2201.04378",
    archivePrefix = "arXiv",
    primaryClass = "gr-qc",
    doi = "10.1142/S0217751X22501275",
    journal = "Int. J. Mod. Phys. A",
    volume = "37",
    number = "18",
    pages = "2250127",
    year = "2022"
}

@article{R26,
    author = "Deb, Biswajit and Deshamukhya, Atri",
    title = "{Constraining Logarithmic $f(R, T)$ Model Using Dark Energy Density Parameter $\Omega_{\Lambda}$ and Hubble parameter $H_0$}",
    eprint = "2207.10610",
    archivePrefix = "arXiv",
    primaryClass = "gr-qc",
    doi = "10.26565/2312-4334-2024-3-02",
    journal = "East Eur. J. Phys.",
    volume = "2024",
    number = "3",
    pages = "21--26",
    year = "2024"
}

@article{R27,
    author = "D'Ambrosio, Fabio and Fell, Shaun D. B. and Heisenberg, Lavinia and Kuhn, Simon",
    title = "{Black holes in $f(Q)$ gravity}",
    eprint = "2109.03174",
    archivePrefix = "arXiv",
    primaryClass = "gr-qc",
    doi = "10.1103/PhysRevD.105.024042",
    journal = "Phys. Rev. D",
    volume = "105",
    number = "2",
    pages = "024042",
    year = "2022"
}

@article{R28,
    author = "Ilyas, M. and Khan, Sobia and Athar, A. R. and Khan, Fawad and Iqbal, Rohna and Alrebdi, Haifa I. and Nisar, Kottakkaran Sooppy and Abdel-Aty, Abdel-Haleem",
    title = "{Non-commutative geometries and wormhole solutions in $f(G)$ gravity}",
    doi = "10.1142/S0219887824503225",
    journal = "Int. J. Geom. Meth. Mod. Phys.",
    volume = "22",
    number = "05",
    pages = "2450322",
    year = "2025"
}

@article{R29,
    author = "Zhang, Xinyi and Chen, Che-Yu and Reyimuaji, Yakefu",
    title = "{Modified gravity models for inflation: In conformity with observations}",
    eprint = "2108.07546",
    archivePrefix = "arXiv",
    primaryClass = "gr-qc",
    doi = "10.1103/PhysRevD.105.043514",
    journal = "Phys. Rev. D",
    volume = "105",
    number = "4",
    pages = "043514",
    year = "2022"
}

@article{R30,
    author = "DeWitt, Bryce S.",
    editor = "Fang, Li-Zhi and Ruffini, R.",
    title = "{Quantum Theory of Gravity. 1. The Canonical Theory}",
    doi = "10.1103/PhysRev.160.1113",
    journal = "Phys. Rev.",
    volume = "160",
    pages = "1113--1148",
    year = "1967"
}

@article{R31,
    author = "Dzhunushaliev, Vladimir and Folomeev, Vladimir and Kleihaus, Burkhard and Kunz, Jutta",
    title = "{Modified gravity from the quantum part of the metric}",
    eprint = "1312.0225",
    archivePrefix = "arXiv",
    primaryClass = "gr-qc",
    doi = "10.1140/epjc/s10052-014-2743-4",
    journal = "Eur. Phys. J. C",
    volume = "74",
    pages = "2743",
    year = "2014"
}

@article{R32,
    author = "Deb, Biswajit and Deshamukhya, Atri",
    title = "{Inflationary models in $f(R,T)$ gravity: Constraints from $Planck$, BICEP/$Keck$, and Atacama}",
    eprint = "2511.06453",
    archivePrefix = "arXiv",
    primaryClass = "gr-qc",
    month = "11",
    year = "2025"
}

@article{R33,
    author = "Deb, Biswajit and Deshamukhya, Atri",
    title = "{Slow-Roll Hilltop Inflation in~$f(\phi ,T)$ Gravity}",
    eprint = "2408.01863",
    archivePrefix = "arXiv",
    primaryClass = "gr-qc",
    doi = "10.1007/978-981-96-4986-0_21",
    journal = "Springer Proc. Phys.",
    volume = "322",
    pages = "137--140",
    year = "2026"
}

@article{R34,
    author = "Yeasmin, Sabina and Deshamukhya, Atri",
    title = "{Embedding warm natural inflation in f({\ensuremath{\phi}})T gravity}",
    eprint = "2408.00595",
    archivePrefix = "arXiv",
    primaryClass = "gr-qc",
    doi = "10.1142/S0217751X24501586",
    journal = "Int. J. Mod. Phys. A",
    volume = "40",
    number = "03",
    pages = "2450158",
    year = "2025"
}

@article{R35,
    author = "Ashmita and Sarkar, Payel and Das, Prasanta Kumar",
    title = "{Slow-roll inflation in the $f(\phi,T)$ gravity theory}",
    doi = "10.1142/s021988782550286x",
    month = "9",
    year = "2025"
}

@article{R36,
    author = "Herrera, Ramon and Rios, Carlos",
    title = "{Reconstructing inflation and reheating in f({\ensuremath{\phi}})T gravity}",
    eprint = "2210.10080",
    archivePrefix = "arXiv",
    primaryClass = "gr-qc",
    doi = "10.1016/j.aop.2023.169484",
    journal = "Annals Phys.",
    volume = "458",
    pages = "169484",
    year = "2023"
}

@article{R37,
    author = "Louis, Thibaut and others",
    collaboration = "Atacama Cosmology Telescope",
    title = "{The Atacama Cosmology Telescope: DR6 power spectra, likelihoods and {\ensuremath{\Lambda}}CDM parameters}",
    eprint = "2503.14452",
    archivePrefix = "arXiv",
    primaryClass = "astro-ph.CO",
    reportNumber = "FERMILAB-PUB-25-0071-PPD",
    doi = "10.1088/1475-7516/2025/11/062",
    journal = "JCAP",
    volume = "11",
    pages = "062",
    year = "2025"
}

@article{R38,
    author = "Calabrese, Erminia and others",
    collaboration = "Atacama Cosmology Telescope",
    title = "{The Atacama Cosmology Telescope: DR6 constraints on extended cosmological models}",
    eprint = "2503.14454",
    archivePrefix = "arXiv",
    primaryClass = "astro-ph.CO",
    reportNumber = "FERMILAB-PUB-25-0157-PPD",
    doi = "10.1088/1475-7516/2025/11/063",
    journal = "JCAP",
    volume = "11",
    pages = "063",
    year = "2025"
}

@article{R39,
    author = "Camphuis, E. and others",
    collaboration = "SPT-3G",
    title = "{SPT-3G D1: CMB temperature and polarization power spectra and cosmology from 2019 and 2020 observations of the SPT-3G main field}",
    eprint = "2506.20707",
    archivePrefix = "arXiv",
    primaryClass = "astro-ph.CO",
    reportNumber = "FERMILAB-PUB-25-0144-PPD",
    doi = "10.1103/7wt3-9v2y",
    journal = "Phys. Rev. D",
    volume = "113",
    number = "8",
    pages = "083504",
    year = "2026"
}

@article{R40,
    author = "Balkenhol, L. and others",
    title = "{Inflation at the End of 2025: Constraints on $r$ and $n_s$ Using the Latest CMB and BAO Data}",
    eprint = "2512.10613",
    archivePrefix = "arXiv",
    primaryClass = "astro-ph.CO",
    month = "12",
    year = "2025"
}

@article{R41,
    author = "Pal, Barun Kumar and Pal, Supratik and Basu, B.",
    title = "{Mutated Hilltop Inflation : A Natural Choice for Early Universe}",
    eprint = "0908.2302",
    archivePrefix = "arXiv",
    primaryClass = "hep-th",
    doi = "10.1088/1475-7516/2010/01/029",
    journal = "JCAP",
    volume = "01",
    pages = "029",
    year = "2010"
}

@article{R42,
    author = "Kumar Pal, Barun",
    title = "{Mutated hilltop inflation revisited}",
    eprint = "1711.00833",
    archivePrefix = "arXiv",
    primaryClass = "gr-qc",
    doi = "10.1140/epjc/s10052-018-5856-3",
    journal = "Eur. Phys. J. C",
    volume = "78",
    number = "5",
    pages = "358",
    year = "2018"
}

@article{R43,
    author = "Gangopadhyay, Mayukh R. and Khan, Hussain Ahmed and Yogesh",
    title = "{A case study of small field inflationary dynamics in the Einstein{\textendash}Gauss{\textendash}Bonnet framework in the light of GW170817}",
    eprint = "2205.15261",
    archivePrefix = "arXiv",
    primaryClass = "astro-ph.CO",
    doi = "10.1016/j.dark.2023.101177",
    journal = "Phys. Dark Univ.",
    volume = "40",
    pages = "101177",
    year = "2023"
}

@article{R44,
    author = "Hounmenou, Samson S. and Salako, Ines G. and Deb, Biswajit and Deshamukhya, Atri",
    title = "{Observational constraints on~$f (T , \mathcal{T})$
~gravity inflationary scenarios from $Planck$, BICEP/$Keck$, and Atacama}",
    doi = "10.1016/j.aop.2026.170462",
    journal = "Annals Phys.",
    volume = "490",
    pages = "170462",
    year = "2026"
}

@article{R45,
    author = "Garriga, Jaume and Mukhanov, Viatcheslav F.",
    title = "{Perturbations in k-inflation}",
    eprint = "hep-th/9904176",
    archivePrefix = "arXiv",
    reportNumber = "UAB-FT-466",
    doi = "10.1016/S0370-2693(99)00602-4",
    journal = "Phys. Lett. B",
    volume = "458",
    pages = "219--225",
    year = "1999"
}

@article{R46,
author = {Vitohekpon, Isaac M. and Salako, Ines G. and Deb, Biswajit and Deshamukhya, Atri},
title = {Inflation in Rastall-Rainbow gravity: confrontation with Planck, BICEP/Keck, and ACT observations},
journal = {International Journal of Modern Physics A},
volume = {0},
number = {ja},
pages = {null},
year = {0},
doi = {10.1142/S0217751X26501034}}

@article{R47,
    author = "Allys, E. and others",
    collaboration = "LiteBIRD",
    title = "{Probing Cosmic Inflation with the LiteBIRD Cosmic Microwave Background Polarization Survey}",
    eprint = "2202.02773",
    archivePrefix = "arXiv",
    primaryClass = "astro-ph.IM",
    doi = "10.1093/ptep/ptac150",
    journal = "PTEP",
    volume = "2023",
    number = "4",
    pages = "042F01",
    year = "2023"
}

@book{R48,
    author = "Abazajian, Kevork N. and others",
    collaboration = "CMB-S4",
    title = "{CMB-S4 Science Book, First Edition}",
    eprint = "1610.02743",
    archivePrefix = "arXiv",
    primaryClass = "astro-ph.CO",
    reportNumber = "FERMILAB-FN-1024-A-AE",
    doi = "10.2172/1352047",
    month = "10",
    year = "2016"
}

@article{A1,
    author = "Ellis, John and Garcia, Marcos A. G. and Olive, Keith A. and Verner, Sarunas",
    title = "{Constraints on attractor models of inflation and reheating from Planck, BICEP/Keck, ACT DR6, and SPT-3G data}",
    eprint = "2510.18656",
    archivePrefix = "arXiv",
    primaryClass = "hep-ph",
    reportNumber = "UMN-TH-4512/25, FTPI-MINN-25/14, KCL-PH-TH/2025-42, CERN-TH-2025-199",
    doi = "10.1103/d35r-7bn8",
    journal = "Phys. Rev. D",
    volume = "113",
    number = "6",
    pages = "063571",
    year = "2026"
}

@article{A2,
    author = "Modak, Tanmoy",
    title = "{R2-Higgs inflation: R3 contribution and preheating after ACT and SPT data}",
    eprint = "2509.02979",
    archivePrefix = "arXiv",
    primaryClass = "hep-ph",
    doi = "10.1103/srpt-jd6s",
    journal = "Phys. Rev. D",
    volume = "112",
    number = "11",
    pages = "115006",
    year = "2025"
}

@article{A3,
    author = "Byrnes, Christian T. and Cort{\^e}s, Marina and Liddle, Andrew R.",
    title = "{Curvaton in light of ACT results}",
    eprint = "2505.09682",
    archivePrefix = "arXiv",
    primaryClass = "astro-ph.CO",
    doi = "10.1103/x43p-zm85",
    journal = "Phys. Rev. D",
    volume = "113",
    number = "6",
    pages = "063568",
    year = "2026"
}

@article{A4,
    author = "Kallosh, Renata and Linde, Andrei and Roest, Diederik",
    title = "{Atacama Cosmology Telescope, South Pole Telescope, and Chaotic Inflation}",
    eprint = "2503.21030",
    archivePrefix = "arXiv",
    primaryClass = "hep-th",
    doi = "10.1103/d6gn-78hn",
    journal = "Phys. Rev. Lett.",
    volume = "135",
    number = "16",
    pages = "161001",
    year = "2025"
}

@article{A5,
    author = "Odintsov, S. D. and Oikonomou, V. K. and Tsyba, Pyotr and Razina, Olga and Rakhatov, Dauren",
    title = "{String-inspired Gauss-Bonnet Gravity Inflation and ACT}",
    eprint = "2604.18861",
    archivePrefix = "arXiv",
    primaryClass = "gr-qc",
    month = "4",
    year = "2026"
}

@article{A6,
    author = "Yuennan, Jureeporn and Atamurotov, Farruh and Capozziello, Salvatore and Channuie, Phongpichit",
    title = "{Constraining $\beta $-exponential inflation with the latest ACT observations}",
    eprint = "2602.17380",
    archivePrefix = "arXiv",
    primaryClass = "gr-qc",
    doi = "10.1140/epjc/s10052-026-15461-1",
    journal = "Eur. Phys. J. C",
    volume = "86",
    number = "3",
    pages = "237",
    year = "2026"
}

@article{A7,
    author = "Aoki, Shuntaro and Otsuka, Hajime and Yanagita, Ryota",
    title = "{Heavy field effects on inflationary models in light of ACT data}",
    eprint = "2509.06739",
    archivePrefix = "arXiv",
    primaryClass = "hep-ph",
    reportNumber = "RIKEN-iTHEMS-Report-25",
    doi = "10.1088/1475-7516/2025/11/088",
    journal = "JCAP",
    volume = "11",
    pages = "088",
    year = "2025"
}

@article{A8,
    author = "Gialamas, Ioannis D. and Katsoulas, Theodoros and Tamvakis, Kyriakos",
    title = "{Keeping the relation between the Starobinsky model and no-scale supergravity ACTive}",
    eprint = "2505.03608",
    archivePrefix = "arXiv",
    primaryClass = "gr-qc",
    doi = "10.1088/1475-7516/2025/09/060",
    journal = "JCAP",
    volume = "09",
    pages = "060",
    year = "2025"
}

@article{A9,
    author = "Barroso Varela, Miguel and Bertolami, Orfeu and Mantziris, Andreas",
    title = "{Inflationary dynamics of non-minimally coupled f(R) matter-curvature theories}",
    eprint = "2509.01532",
    archivePrefix = "arXiv",
    primaryClass = "gr-qc",
    doi = "10.1088/1475-7516/2026/01/061",
    journal = "JCAP",
    volume = "01",
    pages = "061",
    year = "2026"
}

@article{A10,
    author = "Nojiri, Shin'ichi and Odintsov, Sergei and Oikonomou, V. K.",
    title = "{Ghost-free non-local F(R) gravity compatible with ACT}",
    eprint = "2601.07879",
    archivePrefix = "arXiv",
    primaryClass = "gr-qc",
    reportNumber = "KEK-TH-2800, KEK-Cosmo-0407",
    doi = "10.1016/j.physletb.2026.140290",
    journal = "Phys. Lett. B",
    volume = "874",
    pages = "140290",
    year = "2026"
}

@article{A11,
    author = "Odintsov, S. D. and Oikonomou, V. K. and Tsyba, Pyotr and Razina, Olga and Rakhatov, Dauren",
    title = "{String-inspired Gauss-Bonnet Gravity Inflation and ACT}",
    eprint = "2604.18861",
    archivePrefix = "arXiv",
    primaryClass = "gr-qc",
    month = "4",
    year = "2026"
}

@article{A12,
    author = "Kallosh, Renata and Linde, Andrei",
    title = "{New Exponential and Polynomial $ξ$-attractors}",
    eprint = "2605.04415",
    archivePrefix = "arXiv",
    primaryClass = "hep-th",
    month = "5",
    year = "2026"
}

\end{document}